\documentclass[fleqn,12pt,twoside]{article}
\usepackage{espcrc1}

\usepackage{epsfig}
\usepackage[figuresright]{rotating}

\def\NJ{N_{\J}}
\def\J{J/\psi}
\def\Nc{N_c}
\def\Ncbar{N_{\bar c}}
\def\ccbar{c \bar c}
\def\Nccbar{N_{\ccbar}}
\newcommand{\AmS}{{\protect\the\textfont2
  A\kern-.1667em\lower.5ex\hbox{M}\kern-.125emS}}

\title{Quarkonium Formation at High Energy} 

\author{R. L. Thews\address{Department of Physics, University of Arizona,
        Tucson, AZ 85721 USA}\thanks{Supported by U. S. Department of 
Energy Grant DE-FG03-95ER40937}} 
       
\begin{document}

% typeset front matter
\maketitle

\begin{abstract}
The production of quarkonium in heavy ion collisions is studied at RHIC and LHC
energies.  General arguments indicate that, due to initial production of 
multiple quark pairs in each central
collision, the final population of quarkonium may exhibit significant
enhancements over straightforward extrapolation of behavior at SPS energy.
Explicit calculations based on both a statistical hadronization picture and
a kinetic formation mechanism in a deconfined state verify these general
expectations.  Such enhancements will alter the nature of how quarkonium
yields may be used as a signature of deconfinement.
\end{abstract}

\section{GENERIC EXPECTATIONS FOR LARGE $\Nccbar$}

At heavy ion collision energies at RHIC and LHC, the 
initial number of heavy quark pairs produced in each
collision will be qualitatively different than the number produced
at SPS energies.  For example, the number of charm quark pairs is 
expected to be of the order of ten at RHIC and several hundred at LHC 
in the most central collisions \cite{hardprobes1}. Let us attempt to
extract features of $\J$ formation from the initially-produced charm
quarks which are independent of detailed dynamics.  
We consider scenarios in which the formation of $\J$ is allowed to
proceed through any combination of one of the $\Nc$ charm quarks with
one of the $\Ncbar$ anticharm quarks which result from the initial production
of $\Nccbar$ pairs in a central heavy ion collision.  For a given charm quark, 
one expects then that the probability $\cal{P}$ to form a $\J$ is just proportional
to the number of available anticharm quarks relative to the number of
light antiquarks, 
\begin{equation}
{\cal{P}} \propto \Ncbar / N_{\bar u, \bar d, \bar s} \approx \Nccbar / N_{ch},
\end{equation}
where we normalize the number of light antiquarks by the number of 
produced charged hadrons.  
Since this ratio is generally very small, one can simply multiply by the
number of available charm quarks $\Nc$ to obtain the total number of
$\J$ expected in a given event.  

\begin{equation}
\NJ \propto {\Nccbar}^2 / N_{ch},
\end{equation}
where the use of the initial values $\Nccbar = \Nc = \Ncbar$ is again justified
by the relatively small number of bound states formed.
For an ensemble of events, the average number of $\J$ per event is calculated
from the average value of initial charm $<\Nccbar>$, and we neglect fluctuations
in $N_{ch}$.
\begin{equation}
<\J> = \lambda (<\Nccbar>+1)<\Nccbar> / N_{ch},
\end{equation}
where we place all dynamical dependence in the parameter $\lambda$.
[One can extend this formula to the case where $\J$ formation is effective
not over the entire rapidity range $Y_{total}$, but 
only if the quark and antiquark are within the same rapidity interval 
$\Delta y$. In this case one makes the replacement 
$<\Nccbar>+1 \rightarrow <\Nccbar> + Y_{total} / \Delta y$.]

The essential property of this result is that the growth with energy
of the quadratic dependence on total charm \cite{hardprobes1}
is expected to be much stronger than
the corresponding growth of total particle production in heavy ion collisions
\cite{ncharged}.  $\J$ production without this quadratic mechanism is
typically some small energy-independent fraction of total initial charm
production \cite{hardprobes2}, so that we can expect the quadratic
formation to become dominant at high energy.

We show numerical results in Table 1 for these quantities with a
prefactor $\lambda$ of unity.  Estimates for the charm and particle
numbers are very approximate, but serve to show the anticipated trend
with energy.  At SPS, this formation mechanism is most probably 
insignificant.  At RHIC it is comparable with ``normal" formation, while
at LHC one might expect it to be dominant.  Of course, the exact result
will depend on the details of the physics which controls the formation.

\begin{table}[htb]
\caption{Comparison of $\J$ formation variation with energy}
\label{table:1}
\newcommand{\m}{\hphantom{$-$}}
\newcommand{\cc}[1]{\multicolumn{1}{c}{#1}}
\renewcommand{\tabcolsep}{2pc} % enlarge column spacing
\renewcommand{\arraystretch}{1.2} % enlarge line spacing
\begin{tabular}{@{}llll}
\hline
           & \cc{$SPS$} & \cc{$RHIC$} & \cc{$LHC$}  \\
\hline
$\sqrt s$ (GeV)          & \m18 & \m200 & \m5500  \\
$<\Nccbar>$               & \m0.2 & \m10 & \m200 \\
$N_{ch}$                  & \m1350  & \m3250  & \m16500  \\
$<\NJ>$                   & \m0.00018 & \m0.034 & \m2.4 \\
$\NJ^{initial}$           & \m0.0012 & \m0.06 & \m1.2 \\
\hline
\end{tabular}\\[2pt]
\end{table}

\section{STATISTICAL HADRONIZATION MODEL}

This model 
is motivated by the success of attempts to explain the 
relative abundances of light hadrons produced in high energy interactions in
terms of the predictions of a hadron gas in chemical and thermal equilibrium
\cite{thermalfit}.  Such fits, however, are not able to describe the
abundances 
of hadrons containing charm quarks.  This can be understood in terms of the
long time scales required to approach chemical equilibrium for heavy quarks.
However, it is expected that for high energy heavy ion collisions the
initial production of charm quark pairs exceeds the number expected at 
chemical equilibrium as determined by the light hadron abundances.  The
statistical hadronization model \cite{bms1} 
assumes that at hadronization these charm quarks are distributed into
hadrons according to chemical equilibrium, but adjusted by a factor
$\gamma_c$ which accounts for oversaturation of charm.  One power of
this factor multiplies 
a given thermal hadron population for each charm or anticharm
quark contained in the hadron.  Thus the relative abundance of 
$\J$ to that of D mesons, for example, will be enhanced in this model.
The enhancement factor is determined by conservation of charm, again
using the time scale argument to justify neglecting pair production
or annihilation before hadronization.

%\begin{equation}
%\Nccbar = {1\over 2}\gamma_c N_{open} + {\gamma_c}^2 N_{hidden},
%\end{equation}

\begin{equation}
\Nccbar = {1\over 2} \gamma_c N_{open} {{I_1(\gamma_c N_{open})}\over {I_0
(\gamma_c N_{open})}} + {\gamma_c}^2 N_{hidden},
\end{equation}
where $N_{open}$ and $N_{hidden}$ are calculated in the thermal
equilibrium model grand canonical ensemble.
The ratio of Bessel functions is an 
approximate correction factor \cite{canonical} 
which converts to canonical ensemble particle numbers.
This factor will only be significant for non-central collisions
or lower energies in which the absolute number of charm pairs 
is quite small \cite{thermalmany}. 

\begin{figure}[htb]
\begin{minipage}[t]{80mm}
%\framebox[79mm]{\rule[-26mm]{0mm}{52mm}}
\epsfig{width=7.5cm,figure=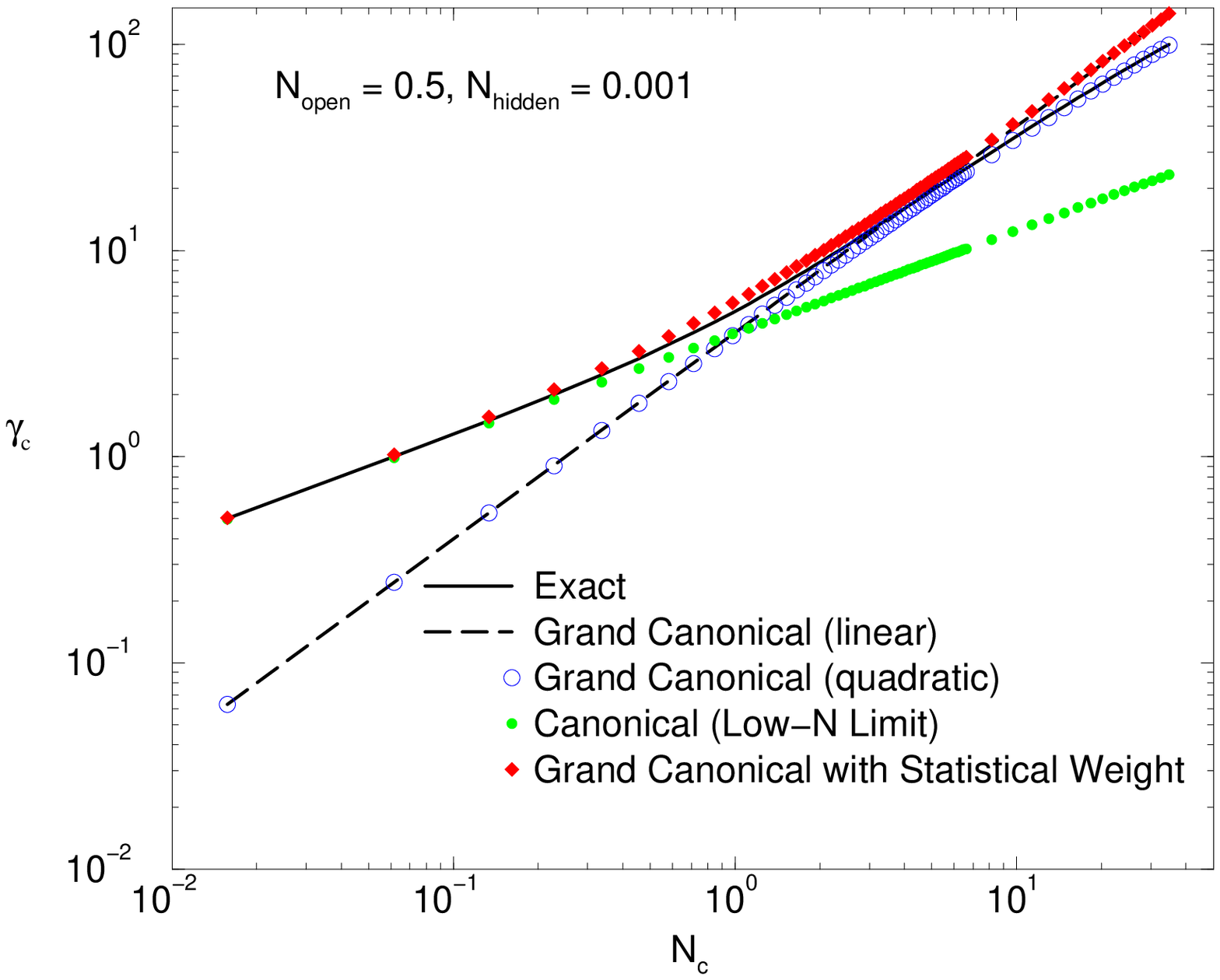}
\caption{\small Relation between charm enhancement factor and total number
of charm quarks for statistical hadronization model.}
\label{gammac}
\end{minipage}
\hspace{\fill}
\begin{minipage}[t]{75mm}
%\framebox[74mm]{\rule[-26mm]{0mm}{52mm}}
\epsfig{width=7.5cm,figure=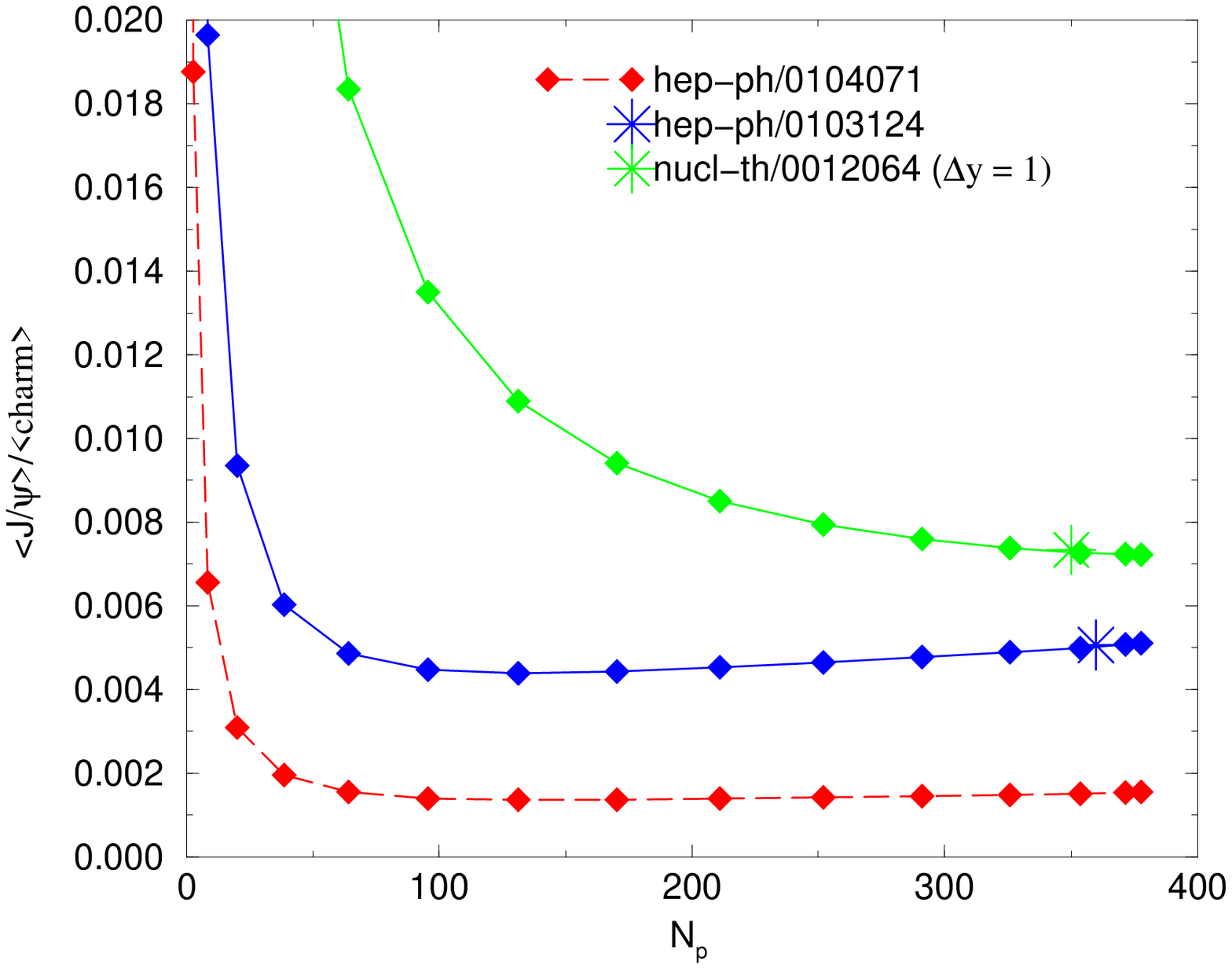}
\caption{\small Ratio $<\J>$ over initial charm at RHIC for several
applications of the statistical
hadronization model.} 
\label{rhicjpsistat}
\end{minipage}
\end{figure}

We show in Figure {\ref{gammac}} calculated $\gamma_c$ values.  At large 
total charm there is linear behavior, which leads to the quadratic
growth of $\J$ formation.  It is interesting to note that the canonical
suppression effect can be very well approximated by using a grand 
canonical formalism, supplemented by the averaging process for 
$\Nccbar^2$ over a distribution of events.

Figure \ref{rhicjpsistat} shows several applications for RHIC conditions.
The centrality dependence is modeled by the number
of nucleon participants, and one sees the change in shape due to the
transition between canonical and grand canonical formalism.  The absolute
magnitudes of $<\J>$/charm are comparable with the initial production 
estimates of a fraction of a percent, indicating that this process may
overwhelm suppression for central collisions at RHIC.
%%%%%%%%%%%%%%%%%%%%%%%%%%%%%%%%%%%%%%%%%%
%%%%%%%%%%%%%%%%%%%%%%%%%%%%%%%%%%%%%%%%

\section{KINETIC FORMATION MODEL}

In this model, we investigate the possibility to form $\J$ directly in
a deconfined medium.  The formation will take advantage of the mobility
of initially-produced charm quarks in a spatial region of
deconfinement, such that all combinations of a charm plus anticharm
are allowed to form a bound state.  
For the purposes of this study, we consider a physical picture
of deconfinement
in which quarkonium is suppressed via collisions with free
gluons \cite{Kha}.  The dominant formation process, 
in which a quark and an antiquark in
a relative color octet state are captured into a color singlet
bound quarkonium state and emit a color octet gluon, 
is simply the
inverse of the breakup reaction.  It is then
an inevitable consequence of this picture of suppression that 
the corresponding formation process must also take place.   

The time evolution if the $\J$ population
is given by the rate equation
\begin{equation}\label{eqkin}
\frac{d\NJ}{d\tau}=
  \lambda_{\mathrm{F}} N_c\, N_{\bar c }[V(\tau)]^{-1} -
    \lambda_{\mathrm{D}} \NJ\, \rho_g\,,
\end{equation}
with $\rho_g$ the number density of gluons.
The reactivity $\lambda$ is
the reaction rate $\langle \sigma v_{\mathrm{rel}} \rangle$
averaged over the momentum distribution of the initial
participants, i.e. $c$ and $\bar c$ for $\lambda_F$ and
$\J$ and $g$ for $\lambda_D$.
The gluon density is determined by the equilibrium value in the
QGP at each temperature.  
The system undergoes a longitudinal isentropic expansion, which fixes
the time-dependence of the volume V($\tau$).
It is evident that the solution of Equation \ref{eqkin} grows quadratically
with initial charm $\Nccbar$, as long as the total $\J \ll \Nccbar$.

\begin{equation}
\NJ(\tau_f) = \epsilon(\tau_f) \times  [\NJ(\tau_0) +
\Nccbar^2 \int_{\tau_0}^{\tau_f}
{\lambda_{\mathrm{F}}\, [V(\tau)\, \epsilon(\tau)]^{-1}\, d\tau}],
\label{eqbeta}
\end{equation}
where $\tau_f$ is the hadronization time determined by the
initial temperature ($T_0$ is a variable parameter) and
final temperature ($T_f$ = 150 MeV ends the deconfining phase).
The function $\epsilon(\tau_f) = 
e^{-\int_{\tau_0}^{\tau_f}{\lambda_{\mathrm{D}}\, \rho_g\,
d\tau}}$ 
would be the suppression factor in this scenario if the
formation mechanism were neglected.

The momentum distribution of the charm quarks is allowed to vary
over a wide range of possibilities.  At one extreme we use
a thermal equilibrium distribution at the QGP temperature.  We also
consider
a distribution unchanged from that introduced in the
initial perturbative QCD processes, plus several intermediate
distributions with decreasing rapidity widths.  The centrality
dependence can also be modeled, using nuclear collision geometry
to set the various spatial scales.  For details, 
see Ref. \cite{prc,qm2001}.

\begin{figure}[htb]
\begin{minipage}[t]{75mm}
%\framebox[79mm]{\rule[-26mm]{0mm}{52mm}}
\epsfig{width=7.5cm,figure=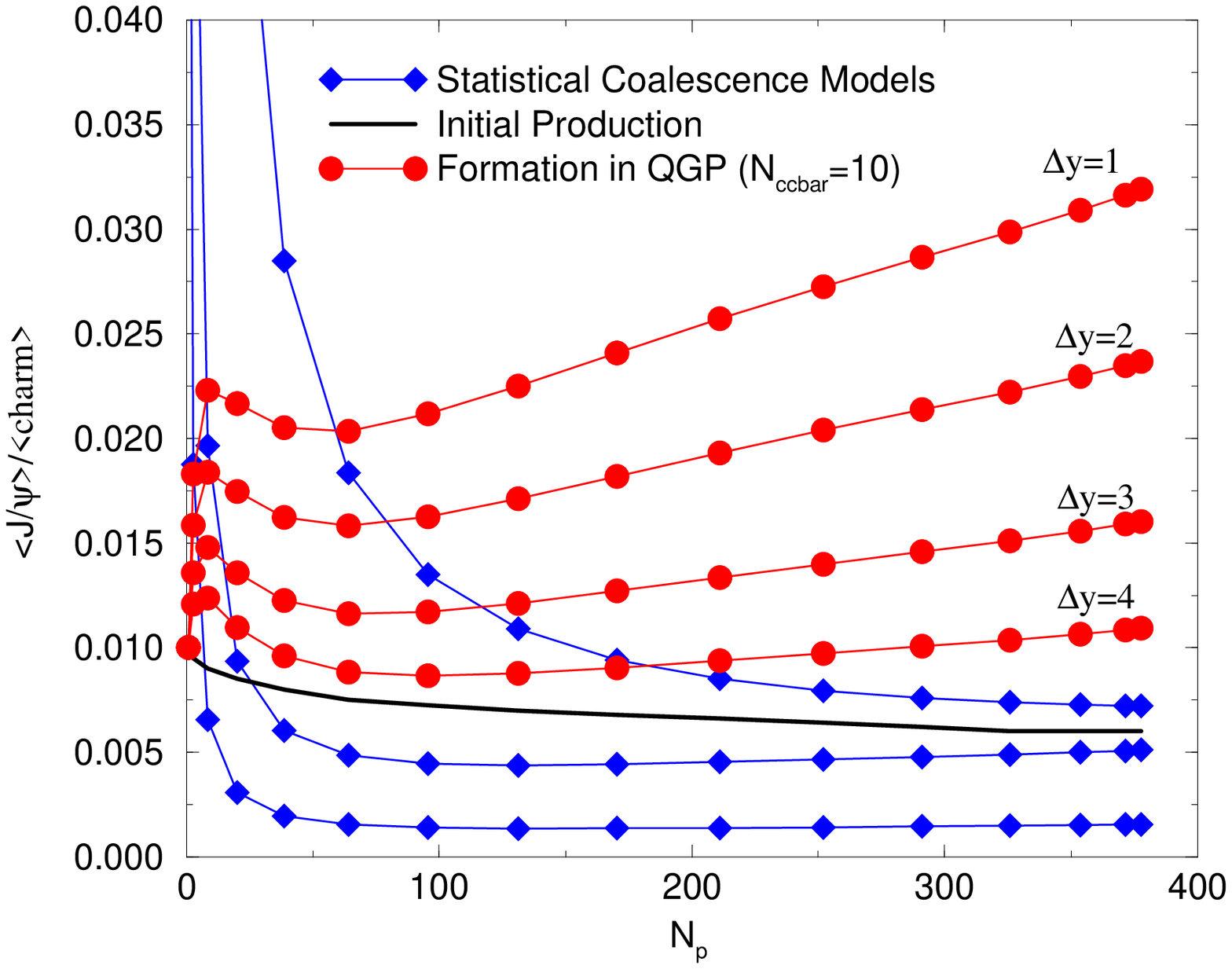}
\caption{\small Predictions of kinetic model 
for $<\J>$ over initial charm at RHIC energy.}
\label{rhicjpsiall}
\end{minipage}
\hspace{\fill}
\begin{minipage}[t]{75mm}
%\framebox[74mm]{\rule[-26mm]{0mm}{52mm}}
\epsfig{width=7.5cm,figure=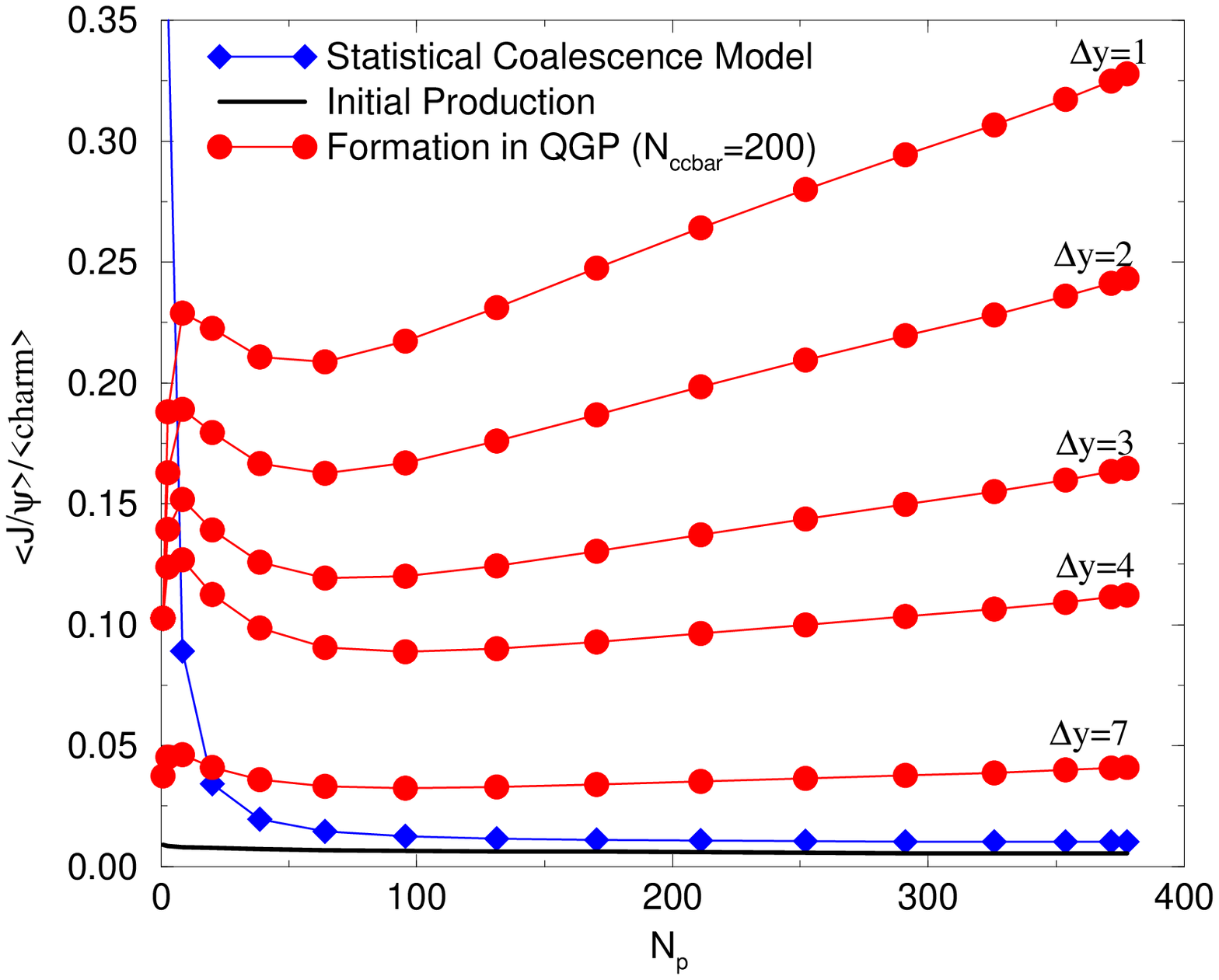}
\caption{\small Predictions of kinetic model for 
$<\J>$ over initial charm at LHC energy.}
\label{lhcjpsiall}
\end{minipage}
\end{figure}

We show the resulting ratios $\NJ$/$\Nccbar$ in Figure \ref{rhicjpsiall}
(for RHIC) and Figure \ref{lhcjpsiall} (for LHC) for a full range of
the initial charm quark rapidity distributions.  Shown for comparison
are some statistical hadronization model results, plus an indication of
the value if only initial (vacuum) production were present.  It is evident
that although the absolute values depend strongly on the charm momentum
distribution, the magnitudes are substantially above that expected if
only initial production plus plasma suppression were present.  The 
centrality dependence is quite striking. Even after normalizing to
initial charm, the ratio rises with increasing centrality.  

%%%%%%%%%%%%%%%%%%%%%%%%%%%%%%%%%%%%%%%%%%
%%%%%%%%%%%%%%%%%%%%%%%%%%%%%%%%%%%%%%%%

\section{SUMMARY}
Expectations based on general grounds for enhanced formation of heavy 
quarkonium in relativistic heavy ion collisions have been verified in
two different models.  In particular, one expects at RHIC and LHC to see
an enhancement in the heavy quarkonium formation rate, even when compared
to unsuppressed production via elementary nucleon-nucleon collisions in
vacuum.  The magnitude of this effect is expected to grow with 
the centrality of the heavy ion collision, just opposite to the predictions
of various suppression scenarios.
The physics bases for these models, however, are
quite distinct.  Their differences should manifest themselves in details of the
magnitudes and centrality dependence.  In this regard, it is essential
to have a simultaneous measurement of open flavor production to serve
as an unambiguous baseline.


\begin{thebibliography}{9}
\bibitem{hardprobes1}  P. L. McGaughey, E. Quack, P. V. Ruuskanen,
R. Vogt, and X.-N. Wang, in ``Hard Processes in Hadronic Interactions'',
           Int. J. Mod. Phys. A10 (1995) 2999.
\bibitem{ncharged}  X.-N. Wang and M. Gyulassy, Phys. Rev.
Lett. 86 (2001) 3496.
\bibitem{hardprobes2}  R. Gavai, D. Kharzeev, H. Satz, G. Schuler,
K. Sridhar, and
R. Vogt, in ``Hard Processes in Hadronic Interactions'',
           Int. J. Mod. Phys. A10 (1995)3043.
\bibitem{thermalfit} See for example P. Braun-Munzinger, D. Magestro, 
K. Redlich, and J. Stachel,
Phys. Lett. B518 (2001) 41, and references therein.
\bibitem{bms1} P. Braun-Munzinger and J. Stachel, Phys. Lett. B490 
(2000) 196.
\bibitem{canonical} J. Cleymans, K. Redlich, and E. Suhonen  
Z. Phys. C51 (1991) 137. 
\bibitem{thermalmany} M.I. Gorenstein, A.P. Kostyuk, H. Stocker, 
and W. Greiner, 
Phys. Lett. B509 (2001) 277; P. Braun-Munzinger and J. Stachel,
Nucl. Phys. A690 (2001) 119.  
\bibitem{Kha}
D. Kharzeev and H. Satz, Phys. Lett. B334 (1994) 155.
\bibitem{prc} R. L. Thews, M. Schroedter, and J. Rafelski,
 Phys. Rev. C63 (2001) 054905. 
\bibitem{qm2001} R. L. Thews and J. Rafelski, $\J$ Production at RHIC
in a QGP, hep-ph/0104025, to be published in proceedings of QM2001.
\end{thebibliography}
\end{document}